# Geochemical discrimination and characteristics of magmatic tectonic settings; a machine learning-based approach

Kenta Ueki[a,b,∗], Hideitsu Hino[c], Tatsu Kuwatani[a,d]

[a]*Department of Solid Earth Geochemistry, Japan Agency for Marine-Earth Science and Technology, 2-15 Natsushima-cho, Yokosuka-city, Kanagawa, 237-0061, Japan*
[b]*Earthquake Research Institute, University of Tokyo, 1-1-1 Yayoi, Bunkyo-ku, Tokyo, 113-0032, Japan*
[c]*Graduate School of Systems and Information Engineering, University of Tsukuba, 1-1-1 Tennodai, Tsukuba, Ibaraki, 305-8573, Japan*
[d]*PRESTO, Japan Science and Technology Agency (JST), 4-1-8 Honcho, Kawaguchi, Saitama, 332-0012, Japan*

**Abstract**

Geochemically discriminating between magmatism in different tectonic settings remains a fundamental part of understanding the processes of magma generation within the Earth's mantle. Here, we present an approach where machine-learning (ML) methods are used for quantitative tectonic discrimination and feature selection using global geochemical datasets containing data for volcanic rocks generated in eight different tectonic settings. This study uses support vector machine, random forest, and sparse multinomial regression (SMR) approaches. All these ML methods with data for 20 elements and 5 isotopic ratios allowed the successful geochemical discrimination between igneous rocks formed in eight different tectonic settings with a discriminant ratio better than 83% for all settings barring oceanic plateaus and back-arc basins. SMR is a particularly powerful and interpretable ML method because it quantitatively identifies geochemical signatures that characterize the tectonic settings of interest and the characteristics of each sample as a probability of the membership of the sample for each setting. We also present the most representative basalt composition for each tectonic setting. The new data provide reference points for future geochemical discussions. Our results indicate that at least 17 elements and isotopic ratios are required to characterize each tectonic setting, suggesting that geochemical tectonic discrimination cannot be achieved using only a small number of elemental compositions and/or isotopic ratios. The results show that volcanic rocks formed in different tectonic settings have unique geochemical signatures, indicating that both volcanic rock geochemistry and magma generation processes are closely connected to the tectonic setting.

∗Corresponding author
*Email address:* `kenta_ueki@jamstec.go.jp` (Kenta Ueki)

## 1. Introduction

The geochemical discrimination of the tectonic setting of magmatic events is one of the most important and useful applications of whole-rock geochemistry (e.g., *Pearce and Cann*, 1973). This approach allows the discrimination of the tectonic setting of a given suite of volcanic rocks using whole-rock geochemical data, including major and trace element and isotopic compositions. These data also allow the identification of similarities and differences between the rock-forming conditions and geological processes that are prevalent in different tectonic settings, including the pressure and temperature conditions during partial melting, phase relations during melting (i.e., degree of partial melting and residual phase assemblages), and source compositions (e.g., anhydrous versus hydrous, variations in the amount of recycled material, and source fertility).

Binary or ternary discrimination diagrams that use only a few elements have been widely used for the geochemical discrimination of the tectonic setting of magmatism since the 1970s (e.g., *Pearce and Cann*, 1973; *Pearce and Norry*, 1979; *Wood*, 1980; *Vermeesch*, 2006a; *Pearce*, 2008). Although these 2D diagrams are useful, easy to use, and are visually easy to understand, recent research has suggested that the use of only a few elements means that these diagrams cannot identify several tectonic settings, and this has cast doubt on the veracity of the results generated by these diagrams (*Snow*, 2006; *Li et al.*, 2015). *Li et al.* (2015) emphasized that effective tectonic discrimination requires numerous elements and recommends the use of multi-element variation diagrams for qualitative tectonic discrimination. In addition to these multi-element diagrams, alternative quantitative geochemical discrimination tools based on decision trees (e.g., *Snow*, 2006) and linear discriminant analysis (e.g., *Agrawal et al.*, 2004; *Vermeesch*, 2006b; *Agrawal et al.*, 2008; *Verma and Armstrong-Altrin*, 2013), have been proposed since the 2000s.

More recent developments in data-analysis techniques, termed machine learning (ML), have also been applied to this problem of geochemical tectonic discrimination. Machine learning is the science of using computers for the automatic detection of patterns in data, and this field has recently developed rapidly in association with advances in computational capabilities (e.g., *Bishop*, 2006). Supervised classification is an ML task where labeled data are considered to represent training data that can be used to determine which discriminants can best classify the dataset in question. One of the most powerful supervised classification techniques is the support vector machine (SVM) (*Cortes and Vapnik*, 1995). The SVM can determine the most appropriate decision high-dimensional compositional plane that can effectively discriminate samples in different classification categories. For example, *Kuwatani et al.* (2014) divided sediments into those related to the 2011 Tohoku-Oki tsunami and non-tsunami sediments using a linear SVM combined with bulk chemical compositional data. More recently, *Petrelli and Perugini* (2016) applied a kernel-based nonlinear SVM to the geochemical tectonic discrimination problem, resulting in the development of a highly accurate technique for the discrimination between different magmatic tectonic settings. However, the amount of geochemical information extracted from the study was limited because the nonlinear SVM does not explicitly output feature vectors, meaning that the relative contribution of individual elements to the discrimination remains unknown. In addition, even if these outputs are provided it is likely that geochemists would find them difficult to utilize because they are both nonlinear and have high dimensionality.

This study introduces two powerful and quantitative ML techniques that are alternatives to the SVM and can be used for the geochemical discrimination of magmas from known tectonic settings. The first of these techniques is the random forest (RF), an ensemble learning method based on decision trees (*Breiman et al.*, 1984), whereas the other approach is sparse multinomial regression (SMR), a sparse variant of the classical linear discrimination method using a relatively small number of elements. These two methods can be used for discrimination as well as for the extraction of the feature quantities that control this discrimination in high-dimensional space. Although the linearity of the SMR technique means this approach has a lesser classification ability than the SVM and RF, we believe it is a particularly powerful approach to geochemical tectonic discrimination because the SMR allows the determination of the geochemical characteristics of tectonic settings as well as the characteristics of individual samples. The careful investigation of feature quantities, such as the relative importance of different elements and the determination of representative bulk compositions for each tectonic setting, also allows the extraction of geochemical information on melting conditions and processes.

This paper initially explains the methodology involved in the ML techniques used in this study before describing the dataset to which the methods are applied. This study uses geochemical data for volcanic rocks from known settings obtained from the PetDB (http://www.earthchem.org/petdb) and GEOROC (http://georoc.mpch-mainz.gwdg.de/georoc/) global geochemical databases. These data are almost same as those used by *Li et al.* (2015) and *Petrelli and Perugini* (2016), enabling the results of our study to be directly compared with theirs. The results of our discrimination analysis are presented with the associated feature quantities that are then geochemically interpreted in the discussion. Machine learning methods such as the SVM and RF are increasingly being used in the Earth sciences (*Cracknell and Reading*, 2013, 2014; *Belgiu and Drăguţ*, 2016; *Petrelli and Perugini*, 2016; *Li et al.*, 2017; *Rouet-Leduc et al.*, 2017), although to our knowledge the SMR has not previously been used with geochemical data. The SMR yields sparse solutions and our results indicate that this technique successfully identified both elements and isotopes that can be effectively used to discriminate between different tectonic settings. We also discuss the differences and limitations of the ML techniques considered here, and provide an overview of the future potential and possible pitfalls of the use of ML in geochemistry.

## 2. Methods

Some elements in igneous rocks occur at concentrations that yield skewed and non-normal compositional distributions (*Ahrens*, 1954). In addition, igneous rocks can have wide variations in trace element concentrations. These large ranges in compositions can introduce undesirable effects when used in ML-based geochemical classification approaches. As such, the geochemical compositional data ($\boldsymbol{x}$) used during this study were normalized using a one-parameter Box–Cox transform (*Box and Cox*, 1964) before the application of ML algorithms, where the parameter ($\lambda$) is determined by the maximum-likelihood-like approach proposed by *Box and Cox* (1964)

$$x^{(\lambda)} = \begin{cases} \dfrac{x^\lambda - 1}{\lambda} & \lambda \neq 0 \\ \log x & \lambda = 0 \end{cases}. \tag{1}$$

The compositional distribution of each element was transformed to be close to a normal distribution using this one-parameter Box–Cox transform. This was followed by the further normalization of each transformed element's compositional data to yield zero mean and unit variance values. The three ML algorithms were then applied to discriminate samples into eight different classes related to tectonic setting. The R source code, including the data normalization methods, and the three ML algorithms are provided in the Supporting Information.

Cross-validation was used to evaluate the discrimination performance of the ML methods. Cross-validation allows the evaluation of the generalization capability (i.e., the prediction capability for unknown data) of the trained model. All of the classification results obtained during this study used a 10-fold cross-validation (*James et al.*, 2014) where the entirety of the dataset is divided into 10 non-overlapping subsets. One of these subsets was retained to test the classification, which was trained by using the union of the remaining nine subsets. This procedure is repeated 10 times, with each repetition changing the test subset to obtain 10 independent estimates of the accuracy of the discrimination approaches. The average of these 10 estimates is reported as the discrimination accuracy of the machine learning algorithms.

### 2.1. Support Vector Machine

A support vector machine (SVM) is a supervised learning model developed for classification problems. The SVM efficiently allows the non-linear and high-dimensional classification of a dataset by defining the decision planes that discriminate between categories. Previous research by *Petrelli and Perugini* (2016) outlined an SVM with a Gaussian kernel function that was used in combination with the compositional dataset outlined above during this study. The kernel parameter was optimized using a 10-fold cross-validation. The SVM method was originally designed for dichotomy problems but can be extended to multi-class discrimination by constructing a one-versus-one classifier for each pair of eight classes. A detailed description of the SVM and its multi-class extension is given by *James et al.* (2014) and *Petrelli and Perugini* (2016).

*2.2. Random Forest*

Random Forest (RF) (*Breiman*, 2001) is a popular and versatile ensemble learning method for discrimination and regression, and has become popular in Earth science research. For example, the RF is widely used with remote sensing data (see the review by *Belgiu and Drăguţ*, 2016). The RF involves thousands of "weak" decision trees (*Breiman et al.*, 1984) that are constructed using randomly selected subsets of samples *and* subsets of features to train the tree. This random selection of samples and features improves the variability of the resulting decision trees, meaning that the RF has a good generalization capability. The RF is also computationally efficient, is known to offer good performance, and can evaluate the importance of different variables. The fact that the trained RF model is composed of thousands of trees that accept subsets of features means that we can evaluate the influence of each of these features on the overall discrimination accuracy. The detailed procedures used to train the RF model and to estimate the importance of different variables are given by *Breiman* (2001).

*2.3. Sparse Multinomial Regression*

Sparse Multinomial Regression (SMR) is a linear classification method that allows the discrimination of different classes as well as feature extraction. the SMR also allows the determination of the importance of features (element concentrations and isotopic ratios) in terms of the discrimination between different classes (tectonic settings) as well as the membership probability that a given sample is within a particular class (tectonic setting). Consequently, the SMR allows the determination of the geochemical characteristics of tectonic settings as well as the characteristics of individual samples.

Multinomial regression is a statistical method that generalizes the logistic regression for multi-class discrimination problems. Both the logistic and multinomial regression are not actually "regression" methods but are instead used for classification. A simplified explanation of the SMR is shown in Figure 1. The multinomial regression uses $\boldsymbol{x} \in \mathbb{R}^p$ as a $p$-dimensional vector with components that correspond to $p$ geochemical elements (i.e., rock compositions) and with $C$ different classes (i.e., tectonic settings). A linear model would usually consider the weighted linear combination of the input vector $w_0+w_1x_1+w_2x_2+\cdots+w_px_p$ with a constant term $w_0$ called *bias*. This notation can be simplified by rewriting $\boldsymbol{x} \leftarrow (1, x_1, x_2, \ldots, x_p)^\top$ and denoting the weighted linear combination with the bias term as

$$\boldsymbol{w}^\top \boldsymbol{x} = w_0 + w_1 x_1 + w_2 x_2 + \cdots + w_p x_p. \qquad (2)$$

The multinomial regression is a linear method that uses a set of projection vectors $\boldsymbol{w}^{(1)}, \boldsymbol{w}^{(2)}, \ldots, \boldsymbol{w}^{(C)} \in \mathbb{R}^{p+1}$ to classify the sample $\boldsymbol{x}$. A sample with a vector representation of $\boldsymbol{x}$ is projected onto $C$ subspaces identified using the projection vectors $\boldsymbol{w}^{(k)}, k = 1, \ldots, C$ as $\boldsymbol{w}^{(1)\top}\boldsymbol{x}, \boldsymbol{w}^{(2)\top}\boldsymbol{x}, \ldots, \boldsymbol{w}^{(C)\top}\boldsymbol{x}$. The probability that the observed data $\boldsymbol{x}$ belongs to the $k$-th class is modeled using

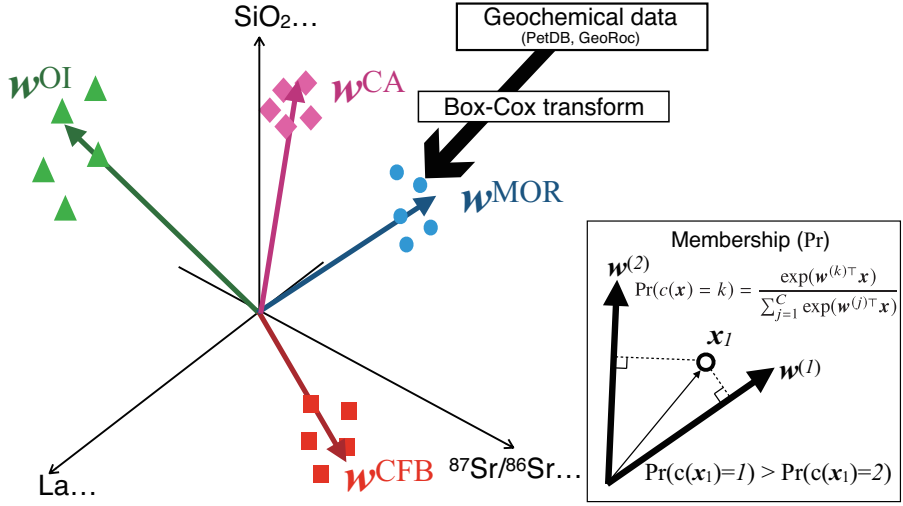

Figure 1: Schematic image of the multinomial regression; see text for a detailed explanation.

$$\Pr(c(\boldsymbol{x}) = k) = \frac{\exp(\boldsymbol{w}^{(k)\top}\boldsymbol{x})}{\sum_{j=1}^{C}\exp(\boldsymbol{w}^{(j)\top}\boldsymbol{x})}, \qquad (3)$$

where $c(\boldsymbol{x})$ is the class assignment function that maps observation $\boldsymbol{x}$ to one of the $C$ classes. The sample $\boldsymbol{x}$ is then classified into the class that gives the maximum probability as: $\mathrm{argmax}_{k\in\{1,2,\ldots,C\}}\Pr(c(\boldsymbol{x}) = k)$.

If we take $n$ samples $\boldsymbol{x}_i, i = 1, 2, \ldots, n$ with each sample $\boldsymbol{x}_i$ belonging to one of $C$ classes and the class label for the sample $\boldsymbol{x}_i$ being denoted by $t(i) \in \{1, 2, \ldots, C\}$ then the projection vectors can be optimized by maximizing the log likelihood function

$$L(\boldsymbol{w}) = \frac{1}{n}\sum_{i=1}^{n}\left\{\boldsymbol{w}^{(t(i))\top}\boldsymbol{x}_i - \log\left(\sum_{j=1}^{C}\boldsymbol{w}^{(j)\top}\exp(\boldsymbol{x}_i)\right)\right\}, \qquad (4)$$

introducing a short notation $\boldsymbol{w} = \{\boldsymbol{w}^{(1)\top}, \ldots, \boldsymbol{w}^{(C)\top}\}^\top$. This study uses the sparse version of multinomial regression (i.e., SMR (*Krishnapuram et al.*, 2005)) for the sake of interpretability and to improve the classification accuracy. Using sparse modeling approach with linear models such as multinomial regressions causes the majority of the elements involved in the projection vectors $\boldsymbol{w}$ to be given values of zero, meaning that only the important elements are included in the model. The sparse projection vector is achieved using

$$\max_{\boldsymbol{w}} \frac{1}{n}\sum_{i=1}^{n}\left\{\boldsymbol{w}^{(t(i))\top}\boldsymbol{x}_i - \log\left(\sum_{j=1}^{C}\boldsymbol{w}^{(j)\top}\exp(\boldsymbol{x}_i)\right)\right\} - \lambda\|\boldsymbol{w}\|_1, \qquad (5)$$

where $\lambda$ is the sparsity-promoting parameter that is determined by 10-fold cross-validation using the training dataset and $\|\boldsymbol{w}\|_1 = \sum_{l=1}^{p+1} |w_l|$ is the $\ell_1$-norm of the vectors. The resulting projection vectors $\boldsymbol{w}^{(j)}$ are expected to be sparse, meaning that most of the $\boldsymbol{w}^{(1)}, \boldsymbol{w}^{(2)}, \ldots, \boldsymbol{w}^{(C)}$ elemental values are exactly zero.

We used the R language (*R Core Team*, 2016), an open source programming language for statistical computing, to implement the ML methods. The R source code used during this analysis is provided in the Supporting Information.

## 3. Dataset

The analysis used whole-rock compositional data for volcanic rocks from known tectonic settings. These data were obtained from the same databases as used by *Petrelli and Perugini* (2016) and used same set of elements and isotopes. The data used are from the PetDB (http://www.earthchem.org/petdb) and GEOROC (http://georoc.mpch-mainz.gwdg.de/georoc/) databases. We only used volcanic rock samples with complete major element data ($SiO_2$, $TiO_2$, $Al_2O_3$, $Fe_2O_3^*$, CaO, MgO, $Na_2O$, $K_2O$), along with selected trace element (Sr, Ba, Rb, Zr, Nb, La, Ce, Nd, Hf, Sm, Gd, Y, Yb, Lu, Ta, Th) and isotopic ($^{206}Pb/^{204}Pb$, $^{207}Pb/^{204}Pb$, $^{208}Pb/^{204}Pb$, $^{87}Sr/^{86}Sr$ and $^{143}Nd/^{144}Nd$) data; if these data were not available then the samples were excluded. Data published after 1990 AD were used to minimize analytical uncertainties, and samples marked as altered and noticeable outliers were excluded from our analysis. Only Quaternary samples were used for arc settings, yielding a total of 2074 samples. Major elements were normalized to 100% anhydrous compositions with total Fe expressed as $Fe_2O_3^*$. The locations of the samples used are shown in Figure 2.

The dataset used in this study is available in the Supporting Information, and salient statistics for these data are given in Table 1. Box plots showing the variations in compositions of volcanic rocks from different tectonic settings are shown in Figure 3. Binary diagrams showing variations in selected element concentrations and isotopic ratios are also shown in Figure 4. These data are also plotted in two trace element binary discrimination diagrams outlined in previous studies (the Th/Yb versus Nb/Yb binary plot and the Th/Yb versus Nb/Yb binary plot; *Pearce and Cann*, 1973; *Pearce and Norry*, 1979; *Pearce*, 2008; *Li et al.*, 2015) in Figure 4. Although some systematic differences between different tectonic settings are present, the compositional ranges of magmas from different tectonic settings clearly overlap in Figures 3 and 4.

An eight-part tectonic setting classification approach was used to be consistent with previous research (*Li et al.*, 2015; *Petrelli and Perugini*, 2016), with the dataset being subdivided into continental arc (CA), island arc (IA), intra-oceanic arc (IOA), back-arc basin (BAB), continental flood (CFB), mid-ocean ridge (MOR), oceanic plateau (OP), and ocean island (OI) settings. We follow the approach of *Li et al.* (2015) in terms of the selection of localities (Table 1 of *Li et al.*, 2015) and as a result have excluded samples from some localities with complex tectonic settings (e.g., the Azores, the Galápagos, Iceland, Chile, and the Mendocino triple junction).

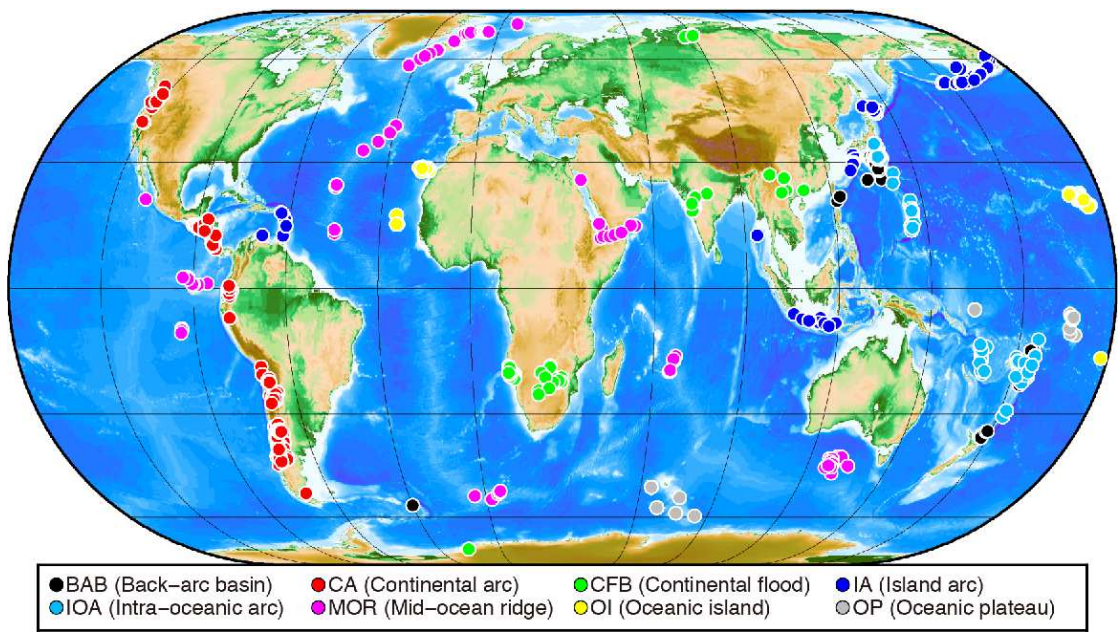

Figure 2: Map showing the location of samples used during this study. This map is based on Terrain-Base (*National Geophysical Data Center/NESDIS/NOAA/U.S. Department of Commerce*, 1995)-derived topographic data.

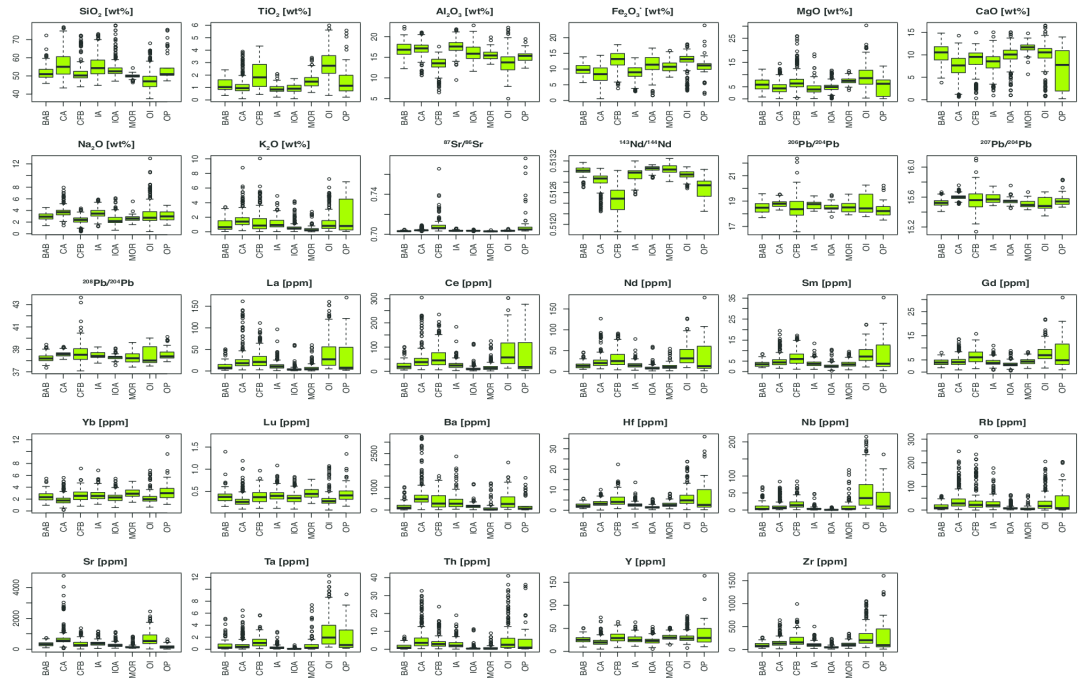

Figure 3: Box plots showing variations in the major and trace element, and isotopic compositions of volcanic rocks from different tectonic settings. The upper and lower sides of each box represent the third and first quartile of the observed values, with the median value shown as the horizontal line dividing the box. The vertical height of the box is thus the interquartile range (IQR). The whiskers (dashed vertical lines) correspond to 1.5×IQR from the first and third quartiles. Observed values that fall beyond these whiskers (marked individually) are generally considered to be outliers.

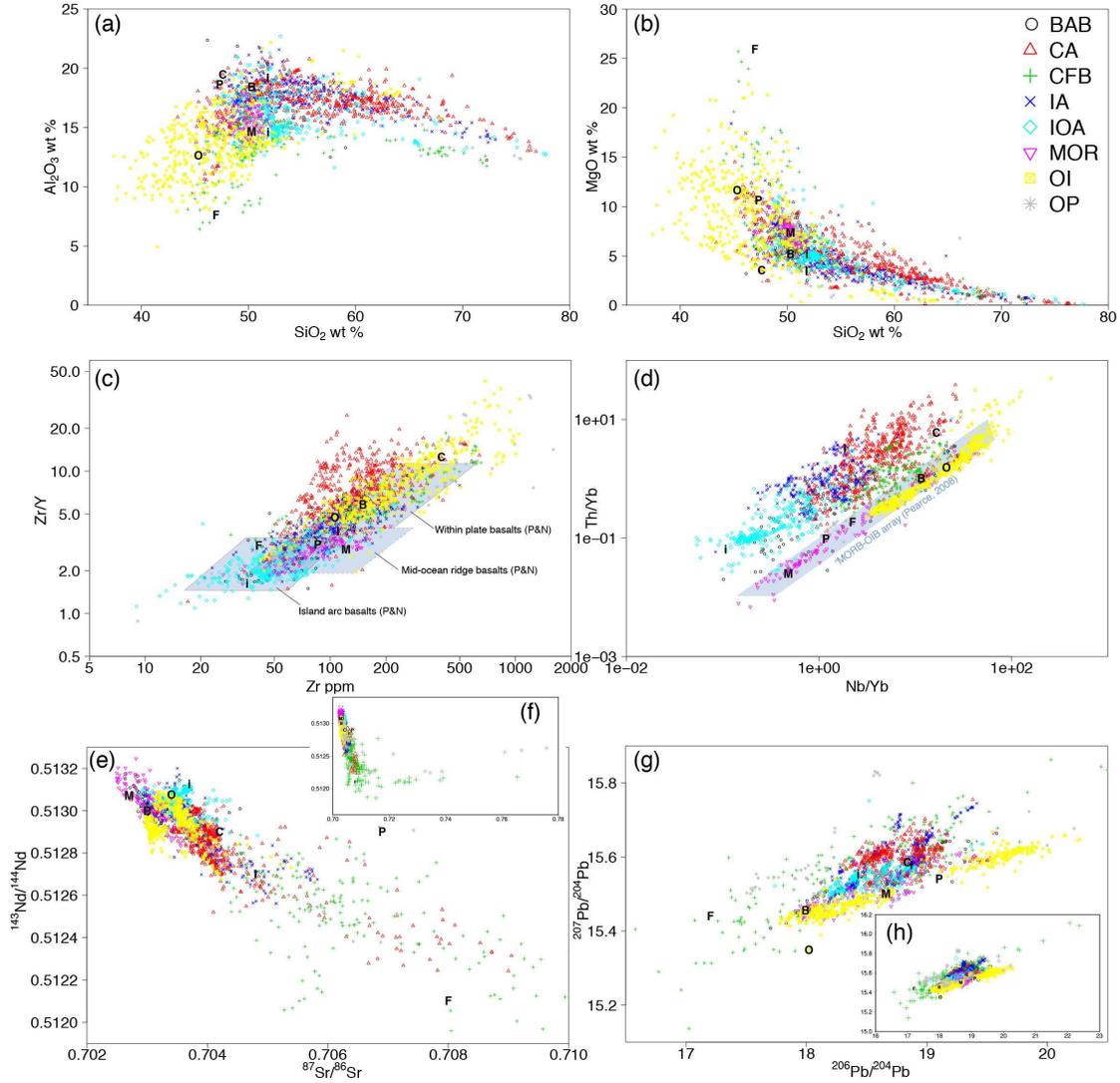

Figure 4: Variations in elemental concentrations and isotopic ratios within the dataset used during this study, shown in (a) $Al_2O_3$ versus $SiO_2$ and (b) MgO versus $SiO_2$ Harker-type diagrams, (c) Zr versus Zr/Y plot (*Pearce and Cann*, 1973; *Pearce and Norry*, 1979; *Li et al.*, 2015) and (d) Th/Yb versus Nb/Yb binary discrimination diagrams (*Pearce*, 2008; *Li et al.*, 2015), and (e) $^{87}Sr/^{86}Sr$ versus $^{143}Nd/^{144}Nd$ and (g) $^{206}Pb/^{204}Pb$ versus $^{207}Pb/^{204}Pb$ isotopic variation diagrams. The full compositional range of the dataset in $^{87}Sr/^{86}Sr$ versus $^{143}Nd/^{144}Nd$ and $^{206}Pb/^{204}Pb$ versus $^{207}Pb/^{204}Pb$ space is shown in the inset figures (Figure 4f and h, respectively). Discrimination fields of *Pearce and Norry* (1979) and MORB–OIB array of *Pearce* (2008) are shown in Figure 4c and Figure 4d, respectively. Abbreviations indicate the representative compositions of each tectonic setting (see text for details) as follows: B, back-arc basin; C, continental arc: F, continental flood; I, island arc; i, intra-oceanic arc; M, mid-ocean ridge; O, oceanic islands; P, oceanic plateau.

Table 1: Data statistics. Table showing the number of samples from each tectonic setting and the arithmetic mean, median, and standard deviation (SD) values for key elements within the dataset for each tectonic setting.

| Setting | Abbreviation | No. of samples | SiO$_2$ [wt%] | | | Total alkalis [ Na$_2$O+K$_2$O wt%] | | | Mg# [ molar MgO/(FeO*+MgO) ratio] | | |
|---|---|---|---|---|---|---|---|---|---|---|---|
| | | | Mean | Median | SD | Mean | Median | SD | Mean | Median | SD |
| Back-arc basin | BAB | 94 | 52.08 | 50.97 | 4.38 | 3.90 | 3.60 | 1.41 | 52.84 | 55.32 | 9.83 |
| Continental arc | CA | 579 | 55.96 | 55.03 | 6.41 | 5.34 | 5.12 | 1.46 | 50.14 | 50.77 | 9.22 |
| Continental flood | CFB | 198 | 52.34 | 50.25 | 6.59 | 3.75 | 3.20 | 1.92 | 48.28 | 48.11 | 14.56 |
| Island arc | IA | 273 | 55.74 | 54.27 | 6.24 | 4.72 | 4.46 | 1.32 | 46.43 | 45.33 | 10.08 |
| Intra oceanic arc | IOA | 338 | 53.83 | 52.52 | 5.17 | 3.06 | 2.62 | 1.24 | 44.36 | 42.63 | 10.36 |
| Mid-Ocean Ridge | MOR | 114 | 49.90 | 50.05 | 1.22 | 3.02 | 2.94 | 0.90 | 57.47 | 57.60 | 6.58 |
| Oceanic island | OI | 434 | 47.44 | 46.99 | 4.94 | 4.54 | 3.51 | 2.82 | 54.17 | 55.67 | 12.35 |
| Oceanic plateau | OP | 45 | 55.12 | 50.96 | 8.27 | 5.37 | 4.76 | 2.89 | 40.63 | 49.59 | 20.82 |

| | Abbreviation | Zr [ppm] | | | Nb [ppm] | | | La [ppm] | | | Y [ppm] | | |
|---|---|---|---|---|---|---|---|---|---|---|---|---|---|
| | | Mean | Median | SD | Mean | Median | SD | Mean | Median | SD | Mean | Median | SD |
| | BAB | 96.33 | 82.75 | 52.52 | 10.89 | 3.33 | 15.99 | 12.07 | 7.67 | 11.74 | 26.04 | 25.10 | 7.39 |
| | CA | 150.50 | 135.00 | 84.12 | 10.79 | 6.50 | 12.60 | 23.11 | 18.00 | 20.89 | 20.71 | 19.56 | 8.66 |
| | CFB | 206.00 | 156.00 | 149.50 | 18.63 | 14.34 | 14.46 | 27.02 | 21.00 | 21.41 | 30.55 | 29.00 | 10.62 |
| | IA | 109.28 | 96.84 | 58.86 | 4.35 | 2.80 | 5.08 | 12.64 | 10.72 | 9.66 | 31.20 | 24.70 | 8.50 |
| | IOA | 55.02 | 46.32 | 31.27 | 1.00 | 0.62 | 1.07 | 4.63 | 3.29 | 5.94 | 23.58 | 22.68 | 8.05 |
| | MOR | 113.52 | 105.25 | 59.54 | 13.55 | 3.40 | 23.13 | 9.15 | 4.38 | 12.25 | 31.37 | 29.91 | 7.86 |
| | OI | 271.50 | 203.00 | 193.31 | 50.94 | 35.00 | 45.26 | 38.24 | 26.95 | 31.14 | 30.09 | 28.00 | 10.66 |
| | OP | 317.14 | 95.60 | 375.69 | 31.33 | 10.40 | 36.89 | 34.81 | 7.56 | 40.62 | 37.29 | 29.20 | 27.25 |

# 4. Results

## 4.1. Support Vector Machine

The classification accuracy of the SVM is presented in confusion matrix form (Figure 5). Confusion matrices contain individual columns that represent instances in a predicted class and rows that represent instances in the actual class. The main diagonal cells within these matrices contain percentage values that indicate the proportion of the data that were correctly assigned, whereas off-diagonal cells contain error rate values. For example, the cell at the intersection of the CFB-row and the BAB-column contains the average (calculated by 10-fold cross-validation) value of the number of error cases where samples formed in CFB settings (i.e., with the true class CFB) have been miss-classified as BAB. Previous research by *Petrelli and Perugini* (2016) indicates that volcanic rocks from different tectonic settings can be successfully discriminated (i.e., with high classification scores) using the SVM. This is exemplified by the fact that all tectonic settings barring BAB have classification scores higher than 91% (Figure 5). This is consistent with the findings of *Petrelli and Perugini* (2016), and our BAB data yielded the lowest classification score (74%) of all of the tectonic settings considered here. A significant proportion of these misclassified BAB (16%) samples were erroneously classified as IOA samples.

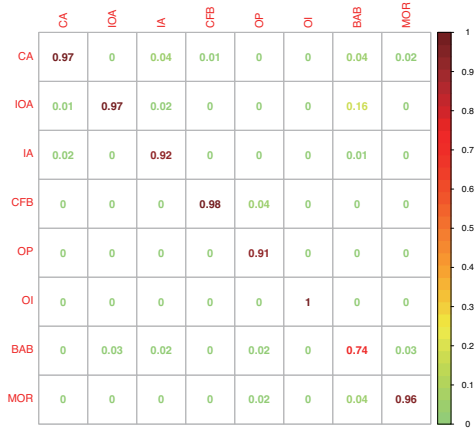

Figure 5: Confusion matrix showing the classification accuracy of the SVM. This confusion matrix contains columns representing instances in a predicted class and rows that represent instances in an actual class. Classification accuracies were calculated using a 10-fold cross-validation.

## 4.2. Random Forest

The confusion matrix obtained from the RF method is shown in Figure 6. The classification scores obtained using this approach are equivalent to, or slightly lower than, those obtained using the SVM. Classification scores higher than 84% were obtained for all tectonic settings with the exception of BAB (49%) and OP (76%), compared with the BAB and OP classification scores by the SVM of 74% and 91%, respectively. The RF resulted in 21% of the BAB samples being misclassified as IOA, 11% of BAB samples being misclassified as MOR, and 15% of OP samples being misclassified as CFB.

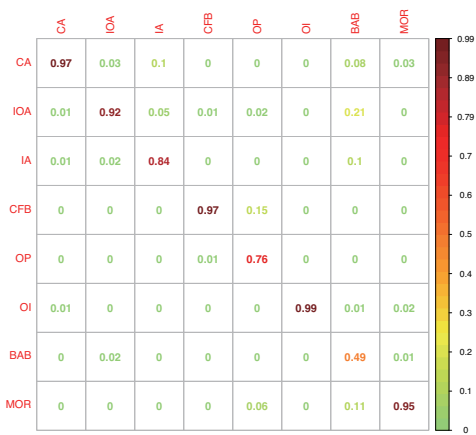

Figure 6: Confusion matrix showing the classification accuracies obtained using the RF.

The RF also allows the determination of the importance of features during discrimination. The importance of different features during the RF classification is shown in Figure 7, revealing that classification accuracies decrease when classifications are performed without individual variables. Highly predictive variables are expected to yield significant decreases as a result of the importance of these variables to the accuracy of the classification. For example, Figure 7 indicates that $^{87}Sr/^{86}Sr$, Nb, $TiO_2$, Sr, and $^{143}Nd/^{144}Nd$ were frequently used by the RF to discriminate between different tectonic settings; i.e., these elements and ratios are particularly important classifiers in the RF. However, any geochemical interpretation is somewhat difficult because the relative importance is derived from the cumulative information within huge numbers of decision trees.

4.3. Sparse Multinomial Regression

The confusion matrix obtained during the SMR analysis is shown in Figure 8. Although the SMR classification scores are the lowest of the three ML approaches used during this study, classification scores higher than 83% were still obtained for all tectonic settings with the exception of BAB and OP. Classification scores of 40% and 76% were obtained for BAB and OP, respectively, with 25% of BAB misclassified as IOA, 14% of BAB misclassified as MOR, and 9% of OP misclassified as CFB.

The fact that the SMR is based on straightforward equations (2 and 3) means that quantitative and interpretable geochemical information can be obtained using this method. The regression coefficients ($w$ in equation 2) provide information on the elements that are important discriminants between magmas from different tectonic settings. In addition, these regression coefficients provide information about the differences in elemental concentrations and isotopic ratio values in volcanic rocks formed in different tectonic settings. Sample membership probabilities for each setting can be obtained from $\Pr(c(\boldsymbol{x}) = k)$ in equation 3, meaning that we can determine the likelihood that a given sample is from a particular tectonic setting using this approach.

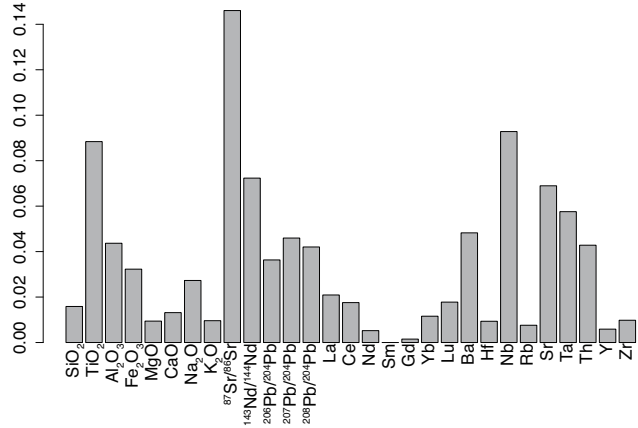

Figure 7: Frequency of appearance of individual parameters during the RF classification.

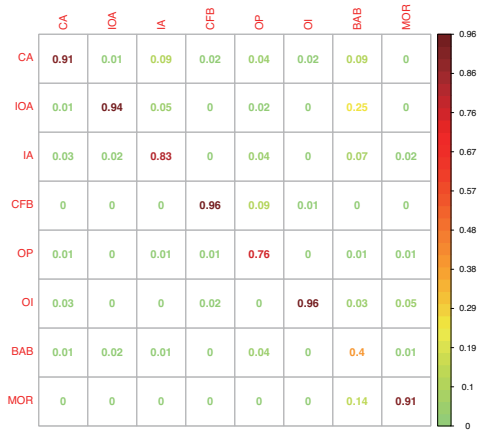

Figure 8: Confusion matrix showing the classification accuracies obtained using the SMR.

Figure 9 shows the regression coefficients $\boldsymbol{w}$ (equation 2) derived by the SMR approach during this study. Multiplying the regression coefficient of a tectonic setting of interest by the composition of a particular sample after the Box–Cox transform (yielding zero mean and unit variance values) as per equations 2 and 3 means that we can obtain the membership probability ($\Pr(c(\boldsymbol{x}) = k)$) of a given sample for a particular tectonic setting (Figure 11). In other words, the absolute values of the regression coefficients represent the importance of each element in discriminating a particular tectonic setting from the other tectonic settings being considered.

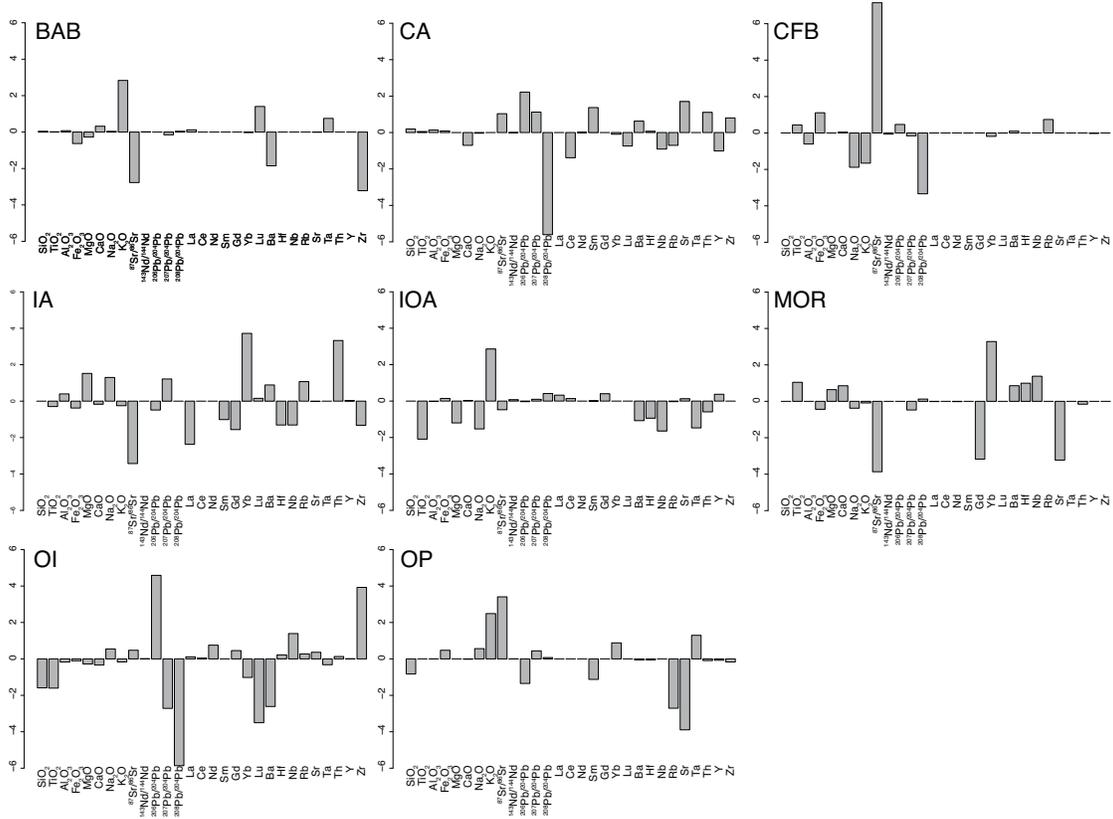

Figure 9: SMR-based regression coefficients $w$ for different tectonic settings, indicating the importance of individual elements in discriminating a given tectonic setting from the remaining seven settings considered here.

The MOR regression coefficients include highly positive loadings for Nb, Hf, Ti, and Yb, and highly negative loadings for $^{87}Sr/^{86}Sr$, Sr, and Gd. A total of 11 out of 29 elements and isotopic ratios yielded zero loadings. The SMR-based regression coefficients for the OI tectonic setting included highly positive loadings for $^{206}Pb/^{204}Pb$, Nb, and Zr, and highly negative loadings for $SiO_2$, $^{207}Pb/^{204}Pb$, $^{208}Pb/^{204}Pb$, $TiO_2$, Yb, and Lu, with only Sm yielding a zero loading. The regression coefficients for the CFB tectonic setting included highly positive loadings for $Fe_2O_3^*$ and $^{87}Sr/^{86}Sr$, highly negative loadings for $Na_2O$, $^{208}Pb/^{204}Pb$, and $K_2O$, and zero loadings for a further 12 elements. The SMR-based regression coefficients for the OP tectonic setting included highly positive loadings for $^{87}Sr/^{86}Sr$, Ta, and $K_2O$, highly negative loadings for $^{206}Pb/^{204}Pb$,

Rb, Sr, and Sm, and zero loadings for a further seven elements. The regression coefficients for the CA tectonic setting included highly positive loadings for $^{87}Sr/^{86}Sr$, $^{206}Pb/^{204}Pb$, $^{207}Pb/^{204}Pb$, Th, Sr, and Sm, highly negative loadings for $^{208}Pb/^{204}Pb$, Nb, Ce, and Y, and zero loadings for a further five elements. The SMR-based regression coefficients for the IA tectonic setting yielded highly positive loadings for MgO, $Na_2O$, $^{207}Pb/^{204}Pb$, Rb, Th, and Yb, highly negative loadings for $^{87}Sr/^{86}Sr$, Nb, La, Zr, Hf, Sm, and Gd, and zero loadings for six further elements or isotopic ratios. The regression coefficients for the IOA tectonic setting included highly positive loadings for $K_2O$, highly negative loadings for MgO, $Na_2O$, Ba, Nb, Ta, Hf, and $TiO_2$, and zero loadings for five elements. The SMR-based regression coefficients for the BAB tectonic setting included highly positive loadings for $K_2O$ and Lu, highly negative loadings for $^{87}Sr/^{86}Sr$, Ba, and Zr, and zero loadings for 10 elements and 1 isotopic ratio ($^{206}Pb/^{204}Pb$). Of the 29 elements and isotopic ratios considered during this study, three major elements ($Fe_2O_3^*$, CaO, and $Na_2O$), one trace element (Ba), and the $^{87}Sr/^{86}Sr$, $^{207}Pb/^{204}Pb$ isotopic ratios had non-zero SMR-based regression coefficients for all tectonic settings.

Binary or ternary diagrams do provide useful information about the data distribution, such as the scattering or clustering of data and differences or similarities between clusters. The data shown in Figure 10 demonstrate how different tectonic settings are discriminated in the SMR-based compositional space. In this figure, samples are plotted based on the most informative feature for a given tectonic setting against all other features that are summarized to yield a single dimension using a weighted projection (equation 2). Figure 10 shows that the tectonic setting of interest can be discriminated by the Box–Cox transformation and the SMR-based projection, whereas the compositional ranges for the other tectonic settings are not discriminated in this diagram.

The SMR-based probabilities ($Pr(c(\boldsymbol{x}) = k)$ in equation 3) of individual samples being assigned to each tectonic setting are shown in Figure 11. The vertical axis in this figure indicates the probability of a sample being associated with the tectonic setting of interest. The data shown in Figure 11 indicate that samples known to have formed in the tectonic setting of interest have a higher probability of being assigned to this tectonic setting than for the other settings. This indicates that the SMR can effectively discriminate tectonic settings of interest from the other settings. In addition, Figure 11 shows a number of intuitive similarities between settings. For example, many BAB samples have high probabilities for MOR and IOA memberships, whereas OI and CFB samples have low probabilities for memberships of other tectonic settings.

The SMR-based tectonic setting membership probabilities (Figure 11) are the basis for the determination of the composition of the most representative basalt within each tectonic setting. Various approaches, including arithmetic means, modes, medians, and logarithmic medians, have been previously used to define "representative compositions" within large datasets. Here, we define the representative sample for each tectonic setting based on the probability that the observed data actually belong to the setting of interest (Figure 11). The basalt ($SiO_2$ = 45–52 wt%) with the highest probability of being formed in a given tectonic setting is here defined as the representative basalt for each individual tectonic setting. The resulting "most representative basalts" are given in Table 2 and are shown in Figures 4 and 12. The representative sample tends to have a distinct, end-member like composition rather than the median- or average-like composition of the setting (Figure 4).

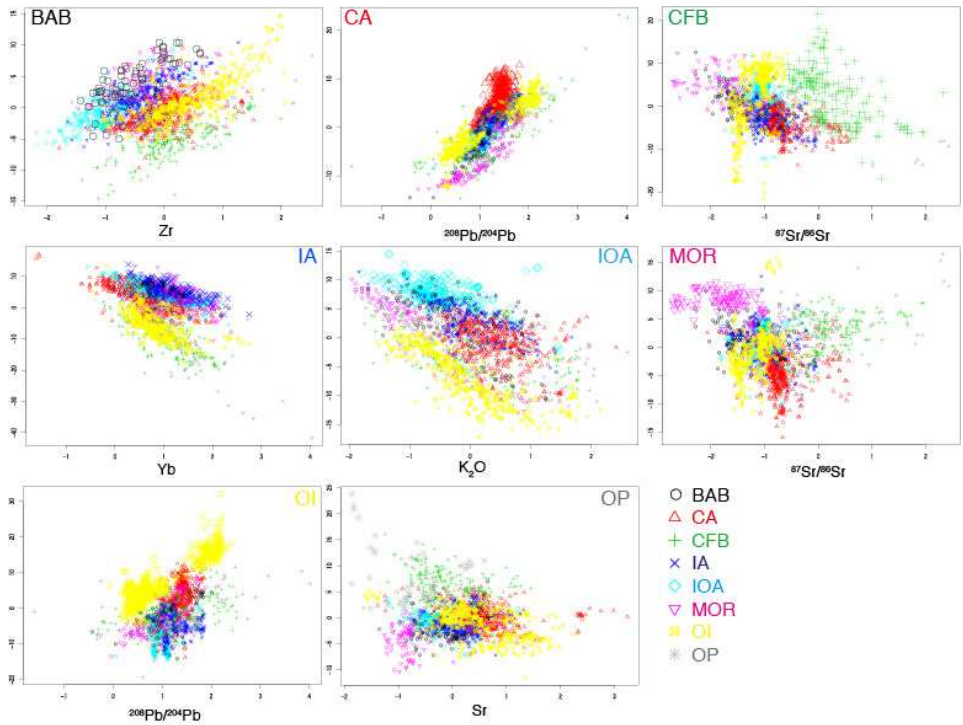

Figure 10: Scatter plot using Box–Cox transformed compositions showing the results of SMR-based discrimination. Samples are plotted on a diagram comparing the most informative feature against all other features summarized to yield one dimension using a weighted projection. Symbols are as in Figure 4.

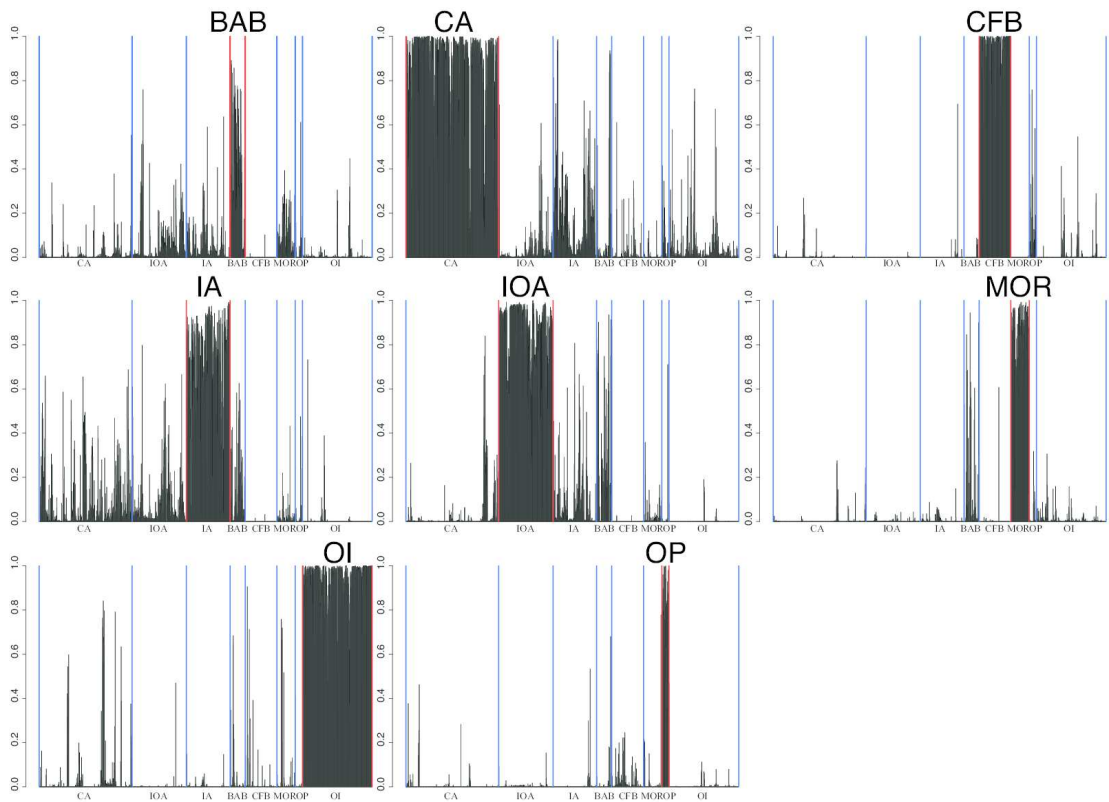

Figure 11: Probabilities of samples for SMR-based membership of each tectonic setting ($\Pr(c(\boldsymbol{x}) = k)$ in equation 3). The vertical axis represents the probability that a given sample was assigned to the tectonic setting of interest. Samples are ordered by the tectonic setting in which they actually formed. The tectonic setting of interest in each panel is indicated by red vertical bars and abbreviations.

Table 2: Representative basalt compositions. Major elements are normalized to 100% anhydrous compositions, with all seven major elements being volatile free and total Fe given as $Fe_2O_3^*$.

| | BAB | CA | CFB | IA | IOA | MOR | OI | OP |
|---|---|---|---|---|---|---|---|---|
| Location | Kinan Seamount Chain | Andean Arc (Sumaco) | Karoo LIP (Plogen) | Sunda Arc (Sangeang Api) | Izu Arc (Oshima) | Southeast Indian Ocean | Hawaii (Oahu) | Ontong Java |
| Reference | *Ishizuka et al. (2009)* | *Chiaradia et al. (2009)* | *Heinonen et al. (2010)* | *Turner et al. (2003)* | *Ishizuka et al. (2015)* | *Kempton et al. (2002)* | *Clague et al. (2006)* | *Shafer et al. (2004)* |
| wt% | | | | | | | | |
| $SiO_2$ | 50.30 | 47.60 | 46.96 | 51.80 | 51.83 | 50.29 | 45.31 | 47.30 |
| $TiO_2$ | 1.65 | 1.41 | 0.49 | 0.80 | 1.18 | 1.78 | 1.98 | 1.26 |
| $Al_2O_3$ | 18.44 | 19.49 | 7.62 | 19.17 | 14.59 | 14.65 | 12.64 | 18.63 |
| $Fe_2O_3^*$ | 8.94 | 10.08 | 13.64 | 8.92 | 15.00 | 11.88 | 13.47 | 18.73 |
| MgO | 5.17 | 3.55 | 25.92 | 3.46 | 5.19 | 7.36 | 11.63 | 10.65 |
| CaO | 10.46 | 8.97 | 4.59 | 8.81 | 10.19 | 11.36 | 11.37 | 0.85 |
| $Na_2O$ | 3.37 | 5.53 | 0.65 | 3.97 | 1.67 | 2.60 | 2.84 | 1.75 |
| $K_2O$ | 1.66 | 3.36 | 0.13 | 3.07 | 0.35 | 0.09 | 0.76 | 0.81 |
| ppm | | | | | | | | |
| La | 18.60 | 100.20 | 4.21 | 33.60 | 1.50 | 3.58 | 26.89 | 8.40 |
| Ce | 37.50 | 189.10 | 10.19 | 64.20 | 4.80 | 11.31 | 51.40 | 17.50 |
| Nd | 19.10 | 80.00 | 7.40 | 29.40 | 4.69 | 10.54 | 26.72 | 10.60 |
| Sm | 4.22 | 12.80 | 2.22 | 6.15 | 1.84 | 3.79 | 6.30 | 3.22 |
| Gd | 4.37 | 8.90 | 2.70 | 0.90 | 2.79 | 4.69 | 5.99 | 4.13 |
| Yb | 2.35 | 2.50 | 1.14 | 2.69 | 2.31 | 3.96 | 1.50 | 3.38 |
| Lu | 0.36 | 0.40 | 0.18 | 0.41 | 0.35 | 0.60 | 0.22 | 0.51 |
| Ba | 218.00 | 2201.00 | 147.00 | 1746.00 | 196.37 | 6.80 | 374.00 | 54.10 |
| Hf | 3.28 | 4.40 | 1.36 | 2.97 | 1.13 | 2.99 | 2.76 | 2.45 |
| Nb | 27.20 | 41.40 | 2.55 | 5.00 | 0.25 | 1.90 | 31.62 | 4.00 |
| Rb | 17.60 | 101.80 | 5.50 | 100.00 | 4.70 | 1.00 | 17.90 | 12.90 |
| Sr | 494.00 | 2646.00 | 173.00 | 1006.00 | 170.88 | 108.00 | 609.00 | 31.70 |
| Ta | 2.06 | 1.90 | 0.15 | 0.35 | 0.02 | 0.12 | 1.76 | 0.23 |
| Th | 2.41 | 14.90 | 0.21 | 8.82 | 0.15 | 0.10 | 2.32 | 0.33 |
| Y | 25.80 | 31.70 | 13.69 | 29.00 | 21.98 | 43.20 | 22.41 | 28.00 |
| Zr | 150.00 | 398.00 | 41.00 | 110.00 | 35.61 | 121.60 | 106.00 | 85.50 |
| Isotopic ratio | | | | | | | | |
| $^{87}Sr/^{86}Sr$ | 0.7030 | 0.7042 | 0.7080 | 0.7048 | 0.7037 | 0.7027 | 0.7034 | 0.7069 |
| $^{143}Nd/^{144}Nd$ | 0.5130 | 0.5129 | 0.5121 | 0.5127 | 0.5131 | 0.5131 | 0.5131 | 0.5129 |
| $^{206}Pb/^{204}Pb$ | 17.99 | 18.84 | 17.20 | 18.87 | 18.43 | 18.66 | 18.02 | 19.10 |
| $^{207}Pb/^{204}Pb$ | 15.46 | 15.59 | 15.44 | 15.58 | 15.55 | 15.50 | 15.35 | 15.54 |
| $^{208}Pb/^{204}Pb$ | 37.88 | 38.57 | 37.93 | 38.87 | 38.35 | 38.22 | 37.51 | 38.76 |

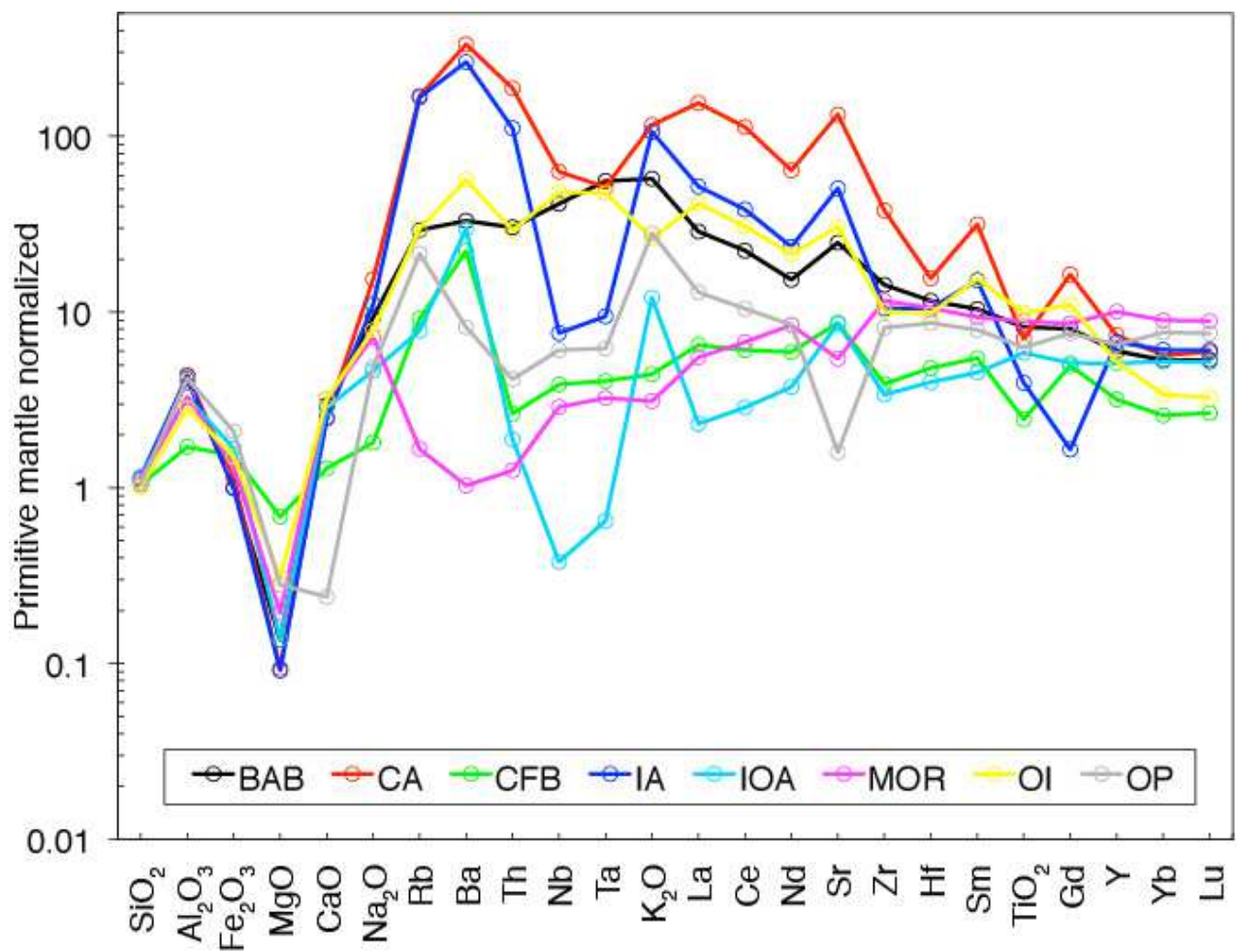

Figure 12: Major and trace element compositions of the representative basalts, as determined using the SMR method and normalized to the primitive mantle composition of *McDonough and Sun* (1995).

## 5. Discussion

### 5.1. Comparison of machine learning methods

The ML analysis undertaken during this study is consistent with the findings of *Petrelli and Perugini* (2016) in that the SVM is a particularly powerful approach for geochemical discrimination. Our analyses indicate that the RF is another powerful technique that could be used for discrimination and feature extraction when addressing geochemical problems. The RF yielded high classification scores that are comparable to those obtained by the SVM. In addition, the RF yields information about the importance of each feature during discrimination. The RF analysis indicates that $^{87}$Sr/$^{86}$Sr, $^{143}$Nd/$^{144}$Nd, Nb, TiO$_2$, and Sr values are commonly important for the geochemical discrimination of tectonic setting (figure 7). These elements and isotopic ratios have often been used as discriminants between different tectonic settings (see *Li et al.* (2015) and references therein), source materials (e.g., *Defant and Drummond*, 1990), and mantle end-members (e.g., *Hofmann*, 2014). However, the quantitative geochemical interpretation of these elements and isotopic ratios obtained by the RF is difficult because these discriminants were only evaluated based on the majority vote of multiple decision trees.

In comparison, the SMR can yield quantitative geochemical interpretations because it is a linear method based on simple equations (e.g., equations 2 and 3). Although the SMR classification scores are relatively low compared with the scores obtained using the SVM and RF, this approach can also yield quantitative knowledge, such as the importance of each feature and the probability of membership in the group of interest for each sample.

*5.2. SMR and feature extraction*

All machine learning methods used in this study can successfully discriminate between volcanic rocks formed in different tectonic settings (Figures 5, 6 and 8), except for BAB. This means that magmas formed in each tectonic setting have unique geochemical signatures. It indicates that the geochemical processes during magma genesis are different in different tectonic setting.

A sparse modeling approach was used to obtain weight values $\boldsymbol{w}$ within a SMR, the majority of which are expected to be zero. This sparse modeling approach allows the recovery of a small number of fundamental signals from a large number of observations. As such, there are cases where a small number of key variables can be identified from within a larger group of several elements or isotopic ratios that exhibit similar chemical properties and trends. For example, when there is a set of several elements that exhibit very good correlation with each other, the geochemical information that they possess can be explained by looking at just one element. In this case, the sparse modeling approach extracts only the most important single element from these several elements. For instance, Gd has strong negative loading whereas Sm has zero loading for MOR, although Sm and Gd are both middle REE (MREE) (Figure 9).

However, at least 17–28 concentrations and isotopic ratios have $\boldsymbol{w}$ values. This means that the magma generation processes that occur in different tectonic settings are essentially complex processes that involve various components and chemical reactions. In addition, our findings are consistent with those of in *Li et al.* (2015) in that the compositional ranges of magmas from different tectonic settings overlap and therefore cannot be discriminated using simple binary or ternary diagrams (figure 4).

Of the tectonic settings investigated during this study, the CFB setting was characterized by the least number of features and could be effectively discriminated using 17 element concentrations and isotopic ratios. In comparison, a total of 28 features were needed to effectively discriminate the OI tectonic setting, 24 for CA and IOA, 23 for IA, 22 for OP, 19 for BAB, and 18 for MOR. This non-sparseness can be understood from a geochemical perspective because the generation and evolution of magmas involves multiple processes such as melting, crystal fractionation, and mixing (e.g., *Ueki and Iwamori*, 2017), and multiple chemical sources such as various mantle components (e.g., *Hofmann*, 1997), subducted slab material, and crustal material (e.g., *Pearce et al.*, 2005). The fact that magmas from a tectonic setting of interest have a wide range in compositions means that a large number of features are required to describe the chemical characteristics of the entire compositional range. For example, basalts and rhyolites within a single arc suite can exhibit distinctive geochemical features, meaning that different discriminators are required for basalts and rhyolites. This in turn means that both of these discriminators need to be represented in the $w$. Non-sparseness related to the OI tectonic setting may also be a result of the wide compositional range of the volcanic rocks that form in this setting (Table 1). Some OI have distinct isotopic compositions that relate to geochemical end-members such as HIMU and EM (e.g., *Stracke*, 2012; *Hofmann*, 2014). This can explain why such a large number of features were derived to cover the vast geochemical variations within the volcanism that occurs in this tectonic setting. The non-sparseness of the subduction zone-related magmas (IA, IOA, CA) may be related to the wide composition of magmas generated in these settings, related to aspects such as across-arc compositional variations between frontal and back-arc volcanoes (e.g., *Kuno*, 1966; *Shibata and Nakamura*, 1997). In addition, arc volcanoes can vary in composition from basaltic to rhyolitic within individual volcanos, as the differentiation of arc-type magmas often occurs within arc crustal settings (e.g., *Ueki and Iwamori*, 2017). In comparison, samples from CFB and MOR settings yield relatively sparse regression coefficients. The sparseness of the CFB data may relate to the distinct compositional range of magmas generated in this setting, especially in terms Sr isotopic compositions (Figure 3). In comparison, the sparseness of the MOR data may relate to the global homogeneity of magmas generated in this type of tectonic setting relative to the other tectonic settings considered here. This means that a small number of features are sufficient to describe the entirety of the geochemical characteristics of MOR.

*5.3. Geochemical characteristics of tectonic settings*

This section focuses on describing the geochemical characteristics of individual tectonic settings using the SMR-based data generated during this study. Our data indicate that the results of our ML-based analysis are consistent with the geochemical characteristics of magmas from different tectonic settings identified during previous geochemical research, such as negative loadings for fluid-immobile elements (i.e., Nb) and positive loadings for fluid-mobile elements (i.e., Rb, Ba, K, and Sr) for magmas from arc-type settings (i.e., CA, IA, and IOA) (*Kogiso et al.*, 1997; *Tatsumi and Kogiso*, 1997; *Schmidt and Poli*, 2014). This means that the data-driven machine learning methods used during this study accurately extracted the geochemical features of different tectonic settings from the geochemical data without any prior knowledge of the geochemical nature of elements or geochemical processes.

*5.3.1. Mid-ocean ridge*

The SMR-based negative loadings for MOR on $^{87}Sr/^{86}Sr$ and $^{207}Pb/^{204}Pb$ are indicative of magmas derived from a depleted source (e.g., *Hofmann*, 1997). A weak positive loading for MOR on $^{208}Pb/^{204}Pb$ may represent its relatively high upper limit of $^{208}Pb/^{204}Pb$ ratio (Figure 3). Both the light rare earth elements (LREE) and the middle REE (MREE) have negative or zero loadings for MOR, meaning that magmas generated in MOR-type settings are depleted in highly incompatible trace elements. The heavy REE (HREE) and the high field strength elements (HFSE) have positive loadings, indicating that MOR magma is generated by low-degree partial melting. In addition, MOR-type magmas have a positive Nb loading, in contrast to arc-type magmas. The SMR-based analysis undertaken during this study is consistent with previous research that determined that MOR is generated by low-degree partial melting of nominally anhydrous depleted mantle material.

*5.3.2. Oceanic island*

The SMR analysis indicates that magmas generated in OI settings have significant Pb isotopic loadings. Figure 10 shows that the OI samples can be divided into two domains based on certain variables (e.g., Pb and Sr isotopes). As discussed above, this may represent the fact that magmas generated in OI settings are derived from regions of the mantle containing various isotopic mantle end-components such as HIMU, EM I, and EM II. In terms of trace elements, OI samples have positive or zero loadings for the LREE and the MREE, whereas the HREE have negative loadings. This means that OI magmas are systematically enriched in the more incompatible trace elements. This is consistent with the results of previous geochemical research that indicate that OI magmas are derived from trace-element-enriched regions of the mantle (e.g., *Hofmann*, 1997). The negative loading on Yb may indicate the presence of residual garnet during mantle melting and the generation of OI type magmas (e.g., *Kimura and Kawabata*, 2014, 2015).

In terms of major elements, both $SiO_2$ and $TiO_2$ have negative loadings, although OI magma tend to have lower $SiO_2$ and higher $TiO_2$ concentrations among the tectonic settings (Figure 3). This finding is possible due to the characteristic of the SMR approach, in which the majority of loadings are expected to be zero. OI magmas exhibit significantly lower $SiO_2$ concentrations than the other settings, and major elements exhibit correlations with $SiO_2$ concentrations (Figure 4). As such, major element characteristics of OI can be mostly characterized solely by the negative loading for $SiO_2$. On the other hand, $TiO_2$ concentrations of OI magmas exhibit the widest compositional range among the tectonic settings (Figure 3). We infer that the negative loading for $TiO_2$ occurs because the samples with lower $TiO_2$ concentrations are the characteristic of OI magmas in the SMR results.

A deep melting source for OI magmas is also consistent with the negative loading for $SiO_2$, because melt $SiO_2$ concentrations decrease with increasing pressure (*Ueki and Iwamori*, 2014).

*5.3.3. Continental flood*

The SVM analysis indicates that CFB samples have high $^{87}Sr/^{86}Sr$ and low $^{208}Pb/^{204}Pb$ values. This isotopic characteristic indicates that CFB magmas are derived from EM-like regions (e.g., *Hofmann*, 1997; *Stracke*, 2012) of the mantle. In addition, Figure 3 indicates that CFB samples are isotopically distinct from magmas generated in other settings. These isotopic characteristics have been attributed to variations in the contribution of fertile sub-continental lithospheric mantle material (*Farmer*, 2014). Our SVM analysis also indicates that CFB samples are characterized by high iron and low $K_2O$ and $Na_2O$ concentrations. *Farmer* (2014) suggested that larger degrees of melting at dry peridotite of shallower depths can generate subalkalic continental flood basalts. *Swanson and Wright* (1981) suggested that the high iron concentrations within the Columbia River basalts indicate that the magmas that formed this CFB were derived from a clinopyroxenite-bearing source region.

*5.3.4. Oceanic plateau*

Both the RF and SMR analyses yielded relatively low classification scores for the OP samples (76%), although the SVM analysis yielded a high classification score of 91%. The majority of misclassified samples were erroneously assigned to the CFB tectonic setting. Previous geochemical research indicates that OP magmas are formed by the partial melting of EM-like enriched mantle material (e.g., *Kerr et al.*, 2014), with the isotopic loadings (positive $^{87}Sr/^{86}Sr$ and negative $^{206}Pb/^{204}Pb$) supporting derivation of OP samples from an EM-like mantle source region. Although both OP and CFB are subdivisions of Large Igneous Provinces (LIPs), our study indicates that OP samples are characterized by EM-like isotopic features whereas CFB samples have HIMU-like signatures. The presence of a strong positive $^{87}Sr/^{86}Sr$ loading is another characteristic of the OP samples analyzed during this study. The high $^{87}Sr/^{86}Sr$ values of OP magmas are associated with either alteration by hydrothermal fluids or by contamination as a result of the assimilation of altered oceanic crustal material (e.g., *Kerr et al.*, 2014). In terms of major element concentrations, low $SiO_2$ concentrations are a key indicator of OP magmas that may (as was the case for OI samples) be indicative of deep-seated partial melting. The negative or zero loadings for the majority of the large ion lithophile elements (LILE) and the REE for the OP samples may be the result of the high-degree partial melting that occurs during the formation of OP magmas.

*5.3.5. Continental arc*

The SMR-based loadings indicate that $^{208}Pb/^{204}Pb$ is the most important discriminator for CA. Compositional range of $^{208}Pb/^{204}Pb$ for CA is the narrowest among the tectonic settings investigated in this study (Figures 3 and 10).

Therefore, it is inferred that $^{208}$Pb/$^{204}$Pb was selected as a effective discriminator that can represent entire CA magmas by the SMR-based regression. In addition to $^{208}$Pb/$^{204}$Pb value, $^{87}$Sr/$^{86}$Sr, $^{206}$Pb/$^{204}$Pb and $^{207}$Pb/$^{204}$Pb values are required to discriminate CA. This means that 4 isotopic data (excluding Nd isotopic values) are required to discriminate CA samples from those formed in other settings. This makes CA magmas distinct from all of the other tectonic settings considered here.

Our SMR result and Figures 4, 3 and 10 indicates that CA magmas are characterized by high $^{87}$Sr/$^{86}$Sr, $^{206}$Pb/$^{204}$Pb, and $^{207}$Pb/$^{204}$Pb values, and intermediate $^{208}$Pb/$^{204}$Pb values. This indicates that CA magmas contain material derived from highly enriched and isotopically evolved (i.e., aged) sources, including subduction-related components (e.g., *Kogiso et al.*, 1997; *Shibata and Nakamura*, 1997; *Pearce et al.*, 2005) and crustal material (e.g., *Kimura and Yoshida*, 2006), both of which characterize CA samples. For example, average upper continental crust falls within the compositional range of MORB in $^{206}$Pb/$^{204}$Pb vs $^{208}$Pb/$^{204}$Pb diagram, whereas it is distinguishable in $^{206}$Pb/$^{204}$Pb vs $^{207}$Pb/$^{204}$Pb diagram (*Hofmann*, 1997). Continental crust is also characterized by its high $^{87}$Sr/$^{86}$Sr (e.g., *Albarède*, 2009). Mixing between mantle-derived magma and crustal-derived felsic magma could thus produce the isotopic signature of CA magma.

In terms of trace elements, CA samples are enriched in Th, Zr, and Sm, all of which are important discriminants between CA magmas and those generated in other subduction-related settings. The Th, Zr, and Sm enriched nature of arc magmas is thought to be a signature of the involvement of crustal melts (*Kimura and Yoshida*, 2006). This indicates that CA magmas are characterized by the involvement of significant amounts of crustal material, unlike the magmas generated in the other tectonic settings considered here. In addition, CA samples are characterized by Sr enrichments and Nb, Ce, and Y depletions. These characteristics may indicate contributions from slab-derived fluids, as Sr is a relatively fluid mobile element, and Nb, Ce, and Y are relatively incompatible in aqueous fluids (e.g., *Kessel et al.*, 2005; *Pearce et al.*, 2005; *Schmidt and Poli*, 2014).

*5.3.6. Intra-oceanic arc*

Unlike CA and IA samples, no significant isotopic loadings were identified for IOA samples during this study. This weak isotopic signature can be explained by the tectonic nature of this magmatism. IOA magmatism occurs in settings where an oceanic plate is subducting beneath another oceanic plate, meaning that IOA settings are underlain by a thin and chemically immature section of the oceanic crust. This means that IOA magmas are minimally influenced by various crustal processes such as the assimilation of mature crustal material as well as crystal fractionation. In addition, the mantle wedge source of IOA magmas is likely to contain minimal amounts of enriched sediment-derived material from the continental crust.

Although IOA magmas experienced the least crustal differentiation among the arc settings, our SMR analysis indicates that IOA magmas contain low concentrations of MgO. This low-MgO feature can also be explained by the tectonic setting of this magmatism. For example, *Tamura et al.* (2016) indicate that the major element compositions of arc magmas from the Izu and Aleutian arcs correlate with the thickness of the overlying crust. *Tamura et al.* (2016) attributed this correlation to the depth of mantle melting, where mantle wedges with thin sections of overlying crust undergo low-pressure melting. The fact that partial melts derived from mantle peridotite material become less olivine-normative with decreasing pressures (e.g., *Ueki and Iwamori*, 2014) means that magmas may also become less olivine-normative in areas with thinner regions of overlying crust. This means that the thinner overlying crust in IOA-type tectonic settings may lead to the generation of low-MgO primary magmas.

In terms of trace elements, IOA samples have negative Ba, Th, Nb, and Ta loadings, and zero or weakly positive loadings for the REE, indicating that these magmas are depleted in the more incompatible trace elements. The majority of the HFSE also have negative loadings for IOA samples. Subduction zone magmas are characterized by HFSE depletions relative to the REE and the LILE (e.g., *Pearce and Peate*, 1995), reflecting subduction-related aqueous fluid processes (e.g., *Münker et al.*, 2004). This means that IOA magmas are generated by the partial melting of a MORB-like region of the depleted mantle that has been influenced by a fluid flux.

*5.3.7. Island arc*

Our SVM analysis indicates that IA samples have features that are intermediate between those of CA and IOA samples. The majority of the LILE have positive loadings for IA samples, whereas the majority of the HFSE have negative loadings, both of which are indicative of the contributions of slab-derived fluids during the genesis of IA magmas. The negative loading for $^{87}Sr/^{86}Sr$ contrasts with the positive loadings for highly enriched settings (CFB, OP, and CA), indicating that this is a useful discriminant for IA samples. In contrast to IOA samples, IA samples have positive loadings for MgO. This relates to the fact that IA settings are characterized by a chain of islands formed by the subduction of an oceanic plate beneath a region of continental crust that is below sea level. As discussed above, the presence of thicker overlying crust within IA settings compared with IOA settings means that the former are characterized by deeper mantle melting that generated higher MgO magmas.

*5.3.8. Back-arc basin*

As noted above and by *Petrelli and Perugini* (2016), BAB samples yield the lowest classification scores of all of the tectonic settings considered during the ML analysis presented here. Most notably, significant numbers of BAB samples were misclassified as IOA and MOR by all of the ML methods. These low classification scores may be related to the complex tectonic configurations of BAB settings. Various back-arc volcanisms, including back-arc spreading centers, back-arc knolls, and seamount chains, are classified as BAB-type volcanism within the PetDB-derived data used during this study. Numerial modeling of the subduction zone has shown that BAB magmatism is triggered by decompression melting by induced upward flow in the mantle wedge (e.g., *Conder et al.*, 2002), as well as slab-related fluid fluxed melting. It has been suggested that the generation processes and source components of BAB magmas change with time during the evolution of the back-arc spreading and slab subduction (e.g., *Ishizuka et al.*, 2009).

The magmas that form some back-arc seamounts, knolls, and spreading centers are known to be derived from enriched regions of the mantle that were affected by subducting slab or metasomatism (e.g., *Ishizuka et al.*, 2010). A significant amount of previous research indicates that back-arc spreading centers also have MORB-like trace element compositions and high concentrations of fluid-mobile elements (e.g., *Gale et al.*, 2013). This means that some BAB samples may have more "arc-like" signatures, whereas others have more "MORB-like" signatures. Coexistence of contrasting magmatic types such as MORB-type magmatism without a subduction influence, and magmas with a signature of contribution of slab-derived fluid have been reported (e.g., *Peate et al.*, 2001; *Ishizuka et al.*, 2009). In addition to the MORB-type and arc-type mantle components, contributions of enriched source mantle have been reported (e.g., *Fretzdorff et al.*, 2002; *Ishizuka et al.*, 2009). The regression coefficients obtained for BAB samples during this study include both arc- and MORB-like signatures, including positive loadings for $K_2O$ and negative loadings for Ba. Our SMR analysis also indicates that BAB samples are isotopically depleted (negative loading for $^{87}Sr/^{86}Sr$).

*5.4. Advantages and limitations of using ML in geochemical research*

The present results indicate that both the RF and the SMR ML analysis can be useful tools to aid the geochemical discrimination of volcanic rocks from known tectonic settings, adding to the SVM ML approach that has previously been used for this type of analysis (*Petrelli and Perugini*, 2016). These ML methods can also contribute to the identification of the tectonic setting of unknown samples. In addition, the RF or the SMR can provide geochemical information such as the identification of key elements and isotopic ratios that enable the discrimination of magmas generated in different tectonic settings. This means that ML methods yield both geochemical information as well as tectonic discrimination when used with large datasets. This study used supervised classification methods and the SMR, SVM, and RF approaches to discriminate tectonic settings. In addition to discrimination, ML techniques can be used to solve problems such as clustering (e.g., *Nakamura et al.*, 2016; *Iwamori et al.*, 2017) to automatically group samples from unlabeled datasets, enable regression approaches that can yield predictive forward models from high-dimensional and complicated data (e.g., *Nakamura et al.*, 2017), and enable dimensional reduction and feature extraction based on basis extraction techniques to find vectors from high-dimensional data (e.g., *Iwamori and Albarède*, 2008; *Ueki and Iwamori*, 2017; *Yoshida et al.*, 2018). In addition, ML approaches can assist in solving many geochemical and geological problems such as the con-struction of thermodynamic models using experimental phase equilibria and the automatic and rapid classification of volcanic tephra.

One limitation is that all ML methods require large datasets. Although significant recent progress has been made in the compilation of geochemical databases, only limited full compositional (i.e., major, trace, and isotopic) geochemical data are available (Table 1). Some geographical sampling bias may also be present in our database (figure 2). In addition, our analysis did not consider the compositional (e.g., *Kimura et al.*, 2017) and thermal evolution (e.g., *Komiya*, 2004) of magmas through geological time, compositional variations within individual tectonic settings (e.g., E-MORB and N-MORB, EM I, EM II, and HIMU, frontal- and rear-arc lavas), variations in fractionation trends such as calc-alkaline and tholeiitic series (*Miyashiro*, 1974), and global geographical systematics (e.g., *Iwamori and Albarède*, 2008; *Kimura et al.*, 2017). Another dataset and associated ML analysis considering age data and sample locations are needed to analyze these variations.

Although ML approaches are certainly useful, caution is required when considering applying ML analysis to geochemical problems. It is important to select appropriate ML methods depending on the situation and the aim of the analysis (*Igarashi et al.*, 2016). In particular, there is still no systematic way for data preprocessing such as data-selecting and standardization as well as hyperparameter tuning during machine learning. These difficulties can be overcome by increasing the number of applications of ML to geochemical problems within collaborations between information scientists and geochemists.

## 6. Conclusions

This study presents the results of a machine-learning-based approach to the geochemical tectonic discrimination of volcanic rocks. We used three different ML methods, namely support vector machine (SVM), random forest (RF) and sparse multinomial regression (SMR) approaches, to whole-rock geochemistry based tectonic discrimination of magmas from eight tectonic settings This study evaluated volcanic rocks from continental arc (CA), island arc (IA), intra-oceanic arc (IOA), back-arc basin (BAB), continental flood (CF), mid-ocean ridge (MOR), oceanic plateau (OP), and ocean island (OI) tectonic settings. All three methods used 20 elements and 3 isotopic ratios to successfully discriminate between these eight tectonic settings with classification accuracies better than 83%, with the exception of samples from OP and BAB settings. This means that magmas from different tectonic settings have unique geochemical signatures, indicating in turn that the geochemical processes during magma genesis are different in different tectonic setting. In addition to geochemical discrimination tools, the RF and SMR approaches also yield geochemical information on individual tectonic settings. Our study indicates that the SMR is a particularly useful and interpretable method for geochemical research because it yields both the quantitative geochemical signatures of tectonic settings and the characteristics of individual samples as well as geochemical discrimination information. As such, although the classification accuracy of the SMR is lower than that of the SVM and RF approaches, the SMR approach can be a powerful tool for geochemical analysis. Geochemical features that allow the discrimination of an individual tectonic setting of interest from other tectonic settings can be quantitatively obtained using the SMR. These SMR-based geochemical signatures can also be contextually geochemically interpreted. In addition, this study objectively defines the composition of representative basalt samples from each tectonic setting using the SMR. These geochemical signatures and representative compositions can be used as references for comparison with newly obtained geochemical data. The result of our SMR analysis indicates that least 17 elements and isotopic ratios are needed to effectively discriminate between volcanism in different tectonic settings, indicating that tectonic discrimination cannot be achieved using only several elements and isotopic ratios. Our results indicate that ML is a highly effective tool in geochemical research that it is particularly useful for the analysis of high-dimensional and huge geochemical datasets.

## 7. acknowledgments

We wish to thank three anonymous reviewers for constructive reviews, and Janne Blichert-Toft for the editorial handling of the manuscript. We thank the Earthquake Research Institute's cooperative research program (Geochemical Data Analysis Using Machine Learning) for useful discussions during this study. Iona McIntosh has improved the manuscript considerably. K.U. is supported by JSPS KAKENHI Grant Numbers JP15H05833 and JP17H02063; H.H. is supported by JSPS KAKENHI Grant numbers JP16K16108, JP25120011, and JST CREST JPMJCR1761; and T.K. is supported by JSPS KAKENHI Grant numbers JP15K20864 and JP2512005, and JST PRESTO Grant Number JP-MJPR1676. Some of the figures in this paper were prepared using Generic Mapping Tools (*Wessel and Smith*, 1998). The data used are from the PetDB (http://www.earthchem.org/petdb) and GEOROC (http://georoc.mpch-mainz.gwdg.de/georoc/) databases. The R source code and dataset used in this study are provided in the Supporting Information.